\documentclass{ws-procs10x7}

\usepackage{graphicx,psfrag}

\catcode`\@=11
\def\@cite#1#2{\unskip\nobreak\relax
    \def\@tempa{$\m@th^{\hbox{\the\scriptfont0 #1}}$}%
    \futurelet\@tempc\@citexx}
\def\@citexx{\ifx.\@tempc\let\@tempd=\@citepunct\else
    \ifx,\@tempc\let\@tempd=\@citepunct\else
    \ifx;\@tempc\let\@tempd=\@citepunct\else
    \let\@tempd=\@tempa\fi\fi\fi\@tempd}
\def\@citepunct{\@tempc\edef\@sf{\spacefactor=\the\spacefactor\relax}\@tempa
    \@sf\@gobble}
\catcode`@=12

\def\OMIT#1{{}}

\def\vereq#1#2{\lower3.5pt\vbox{\baselineskip1.5pt \lineskip1.5pt
\ialign{$#1\hfill##\hfil$\crcr#2\crcr\sim\crcr}}}

\newcommand{\beq}{\begin{equation}}
\newcommand{\eeq}{\end{equation}}
\newcommand{\beqa}{\begin{eqnarray}}
\newcommand{\eeqa}{\end{eqnarray}}

\def\rt{{\tilde{r}}}
\def\zetat{{\tilde{\zeta}}}

\def\Spipi{{S_{\pi\pi}}}

\def\xipipi{{\xi_{\pi\pi}}}
\def\xiKK{{\xi_{K K}}}

\def\zt{\tilde{z}}

\begin{document}

\title{\boldmath CKM matrix from non-leptonic $B$-decays}

\author{A. Salim Safir}

\address{Ludwig-Maximilians-Universit\"at M\"unchen, Department f\"ur Physik,\\
Theresienstra\ss e 37, D-80333 M\"unchen, Germany\\ 
  E-mail: safir@theorie.physik.uni-muenchen.de}

\twocolumn[\maketitle\abstract{

We analyze the impact of the forthcoming CP-violating observables in the 
$B_s\to K^+ K^-$ system, combined with the precise measurement of $\sin2\beta$
,  in the extraction of the CKM matrix. Computing the penguin parameters 
$(r, \theta)$ within QCD factorization yields a precise determination of 
$(\bar\rho, \bar\eta)$, reflected by a weak dependence on $\theta$, which 
is shown to be a second order effect. Using the $SU(3)$-flavour symmetry 
argument and the current $B$-factories data provided by the 
$B_d \to \pi^+ \pi^-$ modes, 
we complement the $B_s \to K^+ K^-$ CP-violating observables in a variety of 
ways, in particular we find that $S_{KK}>0$. Finally, we investigate 
systematically the $SU(3)$-symmetry breaking factor within QCD factorization.
\hspace*{8.cm}\hfill {\footnotesize LMU 12/04}}]

Two body non-leptonic $B$-decays provide an abundant laboratory to access
the CP-violation through $B$-meson decays and to explore
the Unitarity Triangle (UT) of the Cabibbo-Kobayashi-Maskawa (CKM) matrix 
\cite{CKM-C,CKM-KM}. Among them is the gold-plated channel $B_d \to J/\psi K_s$
which is the key mode in  the extraction of the CKM phase $\beta$ with 
marginal hadronic uncertainties. Similarly of capital importance for exploring 
the other two angles, namely $\alpha$ and $\gamma$, is the CP-violation in the
charmless $B$-decays \cite{HQ}, such as $B_d \to \pi\pi, \pi \rho$ 
and their flavour-symmetry related modes. However, in this case the extraction of CKM phases is complicated by the 
so-called penguin pollution.
To deal with that, several analyses have been proposed, mainly based on 
either symmetry argument~\cite{Dunietz:1993rm,Fleischer:1999pa,GL} 
or QCD related approaches \cite{Beneke:2003zv,Bauer:2004tj,Charles:2004jd,Keum:2002cr}.

In this talk, we present the result of \cite{Safir:2004ua}, where we analyzed
 the extraction of CKM parameters from the time-dependent CP-violation 
in $B_s \to K^+ K^-$ decays, which is related to $B_d \to \pi^+ \pi^-$ by 
interchanging all down and strange quarks, i.e.~through the $U$-spin subgroup 
of the $SU(3)$-flavour-symmetry of strong interactions~\cite{Dunietz:1993rm,Fleischer:1999pa}, combined
with the precision observable $\sin 2\beta$. Our estimate of the penguin 
parameters and the $SU(3)$-symmetry breaking factor are carried out in the QCD 
factorization approach~\cite{BBNS3,Beneke:2003zw}.
%

The time-dependent CP-asymmetry in $B_s\to K^+ K^-$ decays
is characterized by two quantities, namely the mixing-induced and 
the direct CP-asymmetries, defined respectively as:
\begin{equation}\label{SandC}
S=\frac{2\, {\rm Im}\xi}{1+|\xi|^2},\qquad
C=\frac{1-|\xi|^2}{1+|\xi|^2},
\end{equation}
where
$\xi=e^{- i\,\phi_s}\,\frac{e^{-i\gamma}+P/T}{e^{+i\gamma}+P/T}$. 
The phase $\phi_s$ denotes the $B_s^0-\bar{B_s^0}$ mixing phase, 
which is almost zero in the Standard Model (SM).
On the other hand, the penguin-to-tree ratio $P/T$, defined above, can be written as:
\begin{equation}\label{ptrphi}
\frac{P}{T}=-\frac{r e^{i\theta}}{\epsilon\sqrt{\bar\rho^2+\bar\eta^2}},
\end{equation}
where the parameters $(r, \theta)$ are pure strong interaction quantities,
$\epsilon\equiv \lambda^2/(1-\lambda^2)$, $\lambda=0.22$ is the Cabibbo 
angle and $(\bar\rho, \bar\eta)$ are the perturbatively improved Wolfenstein
parameters\cite{Wolf}.
Neglecting the very small effects from electroweak penguin contributions 
in our process, one can express the penguin parameter $r\, e^{i\theta}$ 
in the form \cite{BBNS3}
\begin{equation}\label{rqcd}
r\, e^{i\theta}= -
\frac{a^c_4 + r^K_\chi a^c_6 + r_A[b_3+2 b_4]}{
 a_1+a^u_4 + r^K_\chi a^u_6 + r_A[b_1+b_3+2 b_4]},
\end{equation}
where the quantities $r^K_\chi$ and $r_A$ are formally of order 
$\Lambda_{QCD}/m_b$ in the heavy-quark limit. 
A recent analysis in QCD factorization gives~\cite{Safir:2004ua}
\begin{equation}\label{rphi}
r=0.11\pm 0.04, \qquad \theta=0.13\pm 0.31,
\end{equation}
where the error includes an estimate of potentially important power 
corrections.

\begin{figure}[t]
\psfrag{S}{$S_{\pi\pi}$}
\psfrag{etab}{$\bar\eta$}
\begin{center}
\epsfig{file=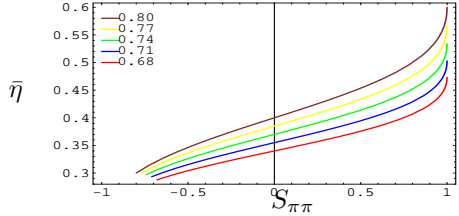,width=6cm,height=3cm}
\end{center}
{\caption{ Lower bound on $\bar\eta$ as a function
of $S_{\pi\pi}$ for various values of $\sin 2\beta$
.  \label{fig:etabound}}}
\end{figure}

The determination of the UT is possible 
by combining the information 
from $S$ with the $B_s^0-\bar{B_s^0}$ mixing phase $\phi_s$ and 
the value of $\sin 2\beta$, which is known with high 
precision from CP-violation measurements in $B\to J/\Psi K_S$.
The angle $\beta$ of the UT is given by $\tau\equiv\cot\beta$.
Using the current world average\cite{HFAG} of $\sin 2\beta$,
implies $\tau=2.26\pm 0.22$.
Given a value of $\tau$, $\bar\rho$ is related to $\bar\eta$
by
\begin{equation}\label{rhotaueta}
\bar\rho = 1-\tau\, \bar\eta.
\end{equation}
Substituting (\ref{rhotaueta}) in (\ref{SandC}), yields \cite{Safir:2004ua}
\begin{eqnarray}\label{etataus}
\bar\eta= {\rm fct}(r, \theta, S, \phi_s, \tau).
\end{eqnarray}
%
%
So far, no approximations have been made and the two expressions in 
(\ref{rhotaueta}) and (\ref{etataus}) are still completely general.
Once the theoretical penguin parameters $r$ and $\theta$ are provided, a 
straightforward determination of the CKM parameters $\bar\eta$ and $\bar\rho$
is obtained from the three observables $\tau$, $\phi_s$ and $S$.
In $B_d\to\pi^+\pi^-$ decays, (\ref{etataus}) leads to model independent 
lower bounds \cite{Buchalla:2003jr} on the CKM parameters $\bar\eta$ for 
$-\sin(2 \beta) \leq S_{\pi\pi} \leq 1$, as shown in Fig. \ref{fig:etabound}. 
They require only the very conservative condition that 
$|\theta|\leq 90^\circ$
.
Moreover, a closer look at the expression in (\ref{etataus}), exhibits a 
rather mild sensitivity of $\bar\eta$ on the strong phase $\theta$.
%
In fact,
the dependence on $\theta$ enters in (\ref{etataus}) only at second order.
Expanding in $\theta$ we obtain\cite{Safir:2004ua}
\begin{equation}\label{etataus0}
\bar\eta= \bar\eta^{(0)} + \bar\eta^{(1)}+....,
\end{equation}
where the leading term $\bar\eta^{(0)}$ is corrected at second order in $\theta$ through $\bar\eta^{(1)}$.
Due to the difficulty in estimating the strong phase, the expression in 
(\ref{etataus0}) is very attractive and consistent with the heavy-quark limit, 
and permits a reasonable extraction of the CKM parameter $\bar\eta$, 
using a quantitative knowledge on the strong phase, as long it is of moderate 
size \cite{Safir:2004ua}.

\begin{figure}
\psfrag{S}{$S_{KK}$}
\psfrag{etab}{$\bar\eta$}
\begin{center}
\epsfig{file=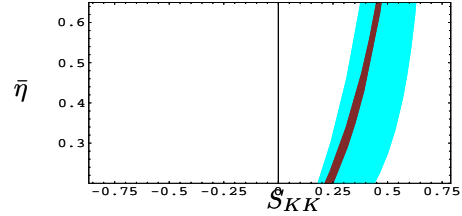,width=6cm,height=3cm}
\end{center}
{\it{\caption{CKM phase $\bar\eta$ as a function of 
$S_{KK}$ within the SM.
The dark (light) band reflects the theoretical uncertainty
in $\theta=0.13\pm 0.31$ 
($r=0.11\pm 0.04$).      
\label{fig:etabspp}}}}
\end{figure}

The determination of $\bar\eta$ as a function of $S$ is depicted
in Fig. \ref{fig:etabspp}. We note that QCD factorization 
prefers positive values of $S$. Furthermore, the sensitivity to 
$\theta$ is less pronounced than for $r$, in extracting $\bar\eta$.
%
Note that, in the absence of penguin contributions, $C_{KK(\pi\pi)}$ would 
vanish and $S_{KK(\pi\pi)}$ would 
provide a clear determination of the phase $\gamma~ (\alpha)$.
However, their presence complicate this determination. In that case, measurement of the CP-violation parameters could be useful in the determination of 
%
$r$ and $\theta$.
Taking \cite{CKMF} $\tau=2.26\pm 0.22 ,\qquad \bar\eta=0.35\pm 0.04$
and the result in (\ref{rphi}), we find from 
(\ref{SandC}) that 
\begin{eqnarray}\nonumber
S= +0.35^{-0.01}_{+0.01}\,(\tau)
         ^{+0.02}_{-0.02}\,(\bar\eta)
         ^{-0.08}_{+0.18}\, (r)
         ^{-0.04}_{-0.00}\,(\theta)\nonumber
\end{eqnarray}
where the dominant uncertainty is due to $r$.

\begin{figure}
\psfrag{rKK}{$r$}
\psfrag{phiKK}{$\theta[rad]$}
\begin{center}
\psfig{file=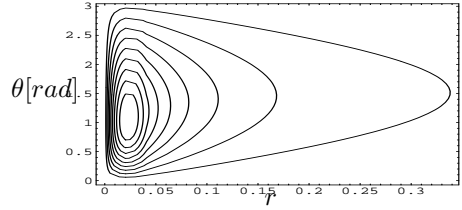,width=6cm,height=3cm}
\end{center}
{\it{\caption{
Contour of constant C in the $(r,\theta)$-plane for fixed $(\bar{\rho},\bar{\eta})=(0.20,0.35)$. 
These contours correspond from right to left, to 
$C$=-0.1,-0.2,-0.3,-0.4,... and -0.9.  
\label{fig:phir}}}}
\end{figure}

On the other hand, the observable $C$ constrains strongly the 
parameters $r$ and $\theta$, as indicated in Fig.~\ref{fig:phir}.
For example, $\theta >(<)~0$ implies that $C<(>)~0$, assuming 
$\bar\eta > 0$. 
However, the corresponding result of the direct CP-asymmetry in 
QCD factorization has a sign opposite to the one found in 
\cite{Charles:2004jd,Buras:2004ub}.

In contrast to $B_d \to \pi^+ \pi^-$ modes, the hadronic quantities $r$ and $\theta$ are less pronounced for the direct CP-violation in $B_s\to K^+ K^-$ than for the former ones~\cite{BS-preparation}. In the SM for fixed $(\bar{\rho},\bar{\eta})=(0.20,0.35)$, a model-independent correlation between $C$ and the hadronic parameters $(r,~\theta)$ implies $r_{max}\approx 0.34$ for $C= \mp 0.1$ and $\theta=\pm\pi/2$. In $B_d \to \pi^+ \pi^-$ decays, this would imply $r'_{max} = 1$ for $C_{\pi\pi}<0.8$ independently of $\theta$~\cite{BS-preparation}. 

As for $B_d\to \pi^+ \pi^-$ cases, a bound on the $B_s\to K^+ K^-$ direct 
CP-violation parameter $C$ as well exists \cite{BS-preparation}: 
\begin{equation}\label{barc}
C_{max}=\frac{-2 \zt\, \sin\theta}{
  \sqrt{(1+\zt^2)^2 -4 \zt^2 \, \cos^2\theta}},
\end{equation}
where the maximum occurs at $\cos\gamma= 2\zt\cos\theta/(1+\zt^2)$, with
$\zt\equiv\bigg|P/T\bigg|$.
Contrary to $B_d\to \pi^+ \pi^-$ modes where $z (\equiv|P/T|_{\pi\pi})\leq 1$, the $B_s\to K^+ K^-$ decay prefers the $\zt \geq 1$ scenario.
%
Then, If $\zt=1$, or equivalently $\rt=R_b$, then $C_{max}\equiv -1$ 
independent
of $\theta$, and no useful upper bound is obtained.
On the other hand, if $\zt > 1$, then $C_{max}$ is maximized for
$\theta=\pi/2$, leading to 
the general bound
\begin{equation}\label{cbound0}
C >- \frac{2\zt}{1+ \zt^2}.
\end{equation}
For the typical value $r \approx 0.15$
, $\zt \approx 7.69$ this implies
$C \geq - 0.26$, which is already a strong constraint on this parameter. The bound on $C$ can be strengthened by using information
on $\theta$, as well as on $\zt$, and employing (\ref{barc}).
Then $\zt \approx  7.69$ and $\theta < 45^\circ~(30^\circ)$ gives 
$C \geq -0.18~ (-0.13)$.

Up to now, our estimate of the CKM parameters has required theoretical 
input on the penguin parameter $(r,\theta)$, which has been 
supplied from QCD factorization.
Another possibility
is to use actual experimental 
informations on the $B_d\to \pi^+ \pi^-$ system in order to estimate 
our penguin parameter $(r,\theta)$, using the $SU(3)$-flavour-symmetry 
argument, on which we will focus below.
For this task, it is convenient, to write in a similar way (\ref{etataus0}) 
for the $B_d\to \pi^+ \pi^-$ channel.

Since the decays $B_d\to \pi^+ \pi^-$ and $B_s\to K^+ K^-$ are related to each other by interchanging all strange and down quarks, the $SU(3)$-flavour-symmetry of strong interactions implies:
\begin{eqnarray}
r=r',\qquad \theta=\theta'.\label{rSU3}
\end{eqnarray}

Assuming that the $B_s^0-\bar{B_s^0}$ mixing phase $\phi_s$ is negligibly 
small, the weak dependence on the strong phase in $\bar\eta$ and considering 
the $SU(3)$-symmetry-breaking effects, we get~\cite{Safir:2004ua}
\begin{eqnarray}\label{rtlSU3}
r'=
\frac{
\xipipi
+
\xiKK      
}
{
      \zetat_{SU3} 
\xiKK
-
\xipipi
},
\end{eqnarray}
where $\xi_{K K,\pi\pi}$ are given in \cite{Safir:2004ua},
$\zetat_{SU3}=\zeta_{SU3}/\epsilon$ and as an {\it educated guess} for our 
analyses, we choose:
\begin{eqnarray}\label{fa-SU3}
\zeta_{SU3}\equiv\frac{r}{r'}=1\pm 0.3, 
\end{eqnarray}
as our $SU(3)$-symmetry-breaking estimate in relating the hadronic physics 
of our corresponding modes.
\begin{figure}[t]
\hspace{-2.cm}
\psfrag{r}{$r'$}
\psfrag{SKK}{$S_{K K}$}
\begin{center}
\epsfig{file=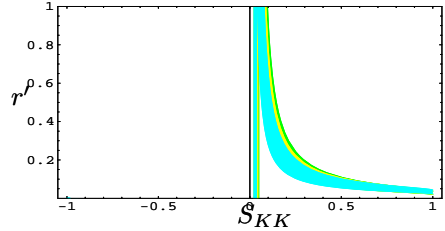,width=6cm,height=3cm}
\end{center}
{\it{\caption{The dependence of $r'$ on $S_{K K}$ fixed through  $\Spipi$ 
and $\zeta_{SU(3)}$. The band reflects the theoretical estimate on the  
$\zeta_{SU(3)}$ for the corresponding $\Spipi=-0.8, -0.5 $ and $-0.1$ 
(from bottom to top). 
\label{fig:rSKK}}}}
\end{figure}

In Fig.~\ref{fig:rSKK}, we plotted the dependence of $r'$ on $S_{K K}$, for 
various values of $S_{\pi \pi}$, 
using our $SU(3)$-symmetry-breaking estimate defined in (\ref{fa-SU3}). 
We observed that constraining our penguin parameter $r'$ (and hence $r$) to 
be positive implies a positive value of $S_{K K}$, in agreement with the 
result obtained in \cite{Buras:2004ub,Charles:2004jd}.


The analyses described above require mainly theoretical input on the $SU(3)$-symmetry-breaking effects $\zeta_{SU(3)}$, which is badly established at present~\cite{Khodjamirian:2003xk}. 
Therefore, we have relied on its generic value, namely $\sim 30\%$, to perform our analyses.
To reinforce the validity of this approximation, it is important to test  
this estimation within the QCD factorization framework, where 
the $SU(3)$-symmetry-breaking effects do enter in several ways \cite{Beneke:2003zw}. In agreement with the result obtained in~\cite{Beneke:2003zw},
we found~\cite{Safir:2004ua} 
\begin{eqnarray}\label{QCDFSU3-f}
\zeta^{\rm{QCDF}}_{SU(3)}=  1.03 \pm  0.09.
\end{eqnarray}
with the dominant error due of the annihilation contributions and the 
Gegenbauer moments of the kaon meson wave function among the remaining input 
parameters. However this error can be large up to $ 30\%$ in magnitude 
assuming non-universality of hard spectator scattering and weak annihilation 
terms (see \cite{Safir:2004ua,Beneke:2003zw} for further details).

In Conclusion, we surveyed the impact of the forthcoming measurements of the 
time-dependent CP-asymmetry parameters in $B_s\to K^+K^-$ decays on the 
extraction of weak phases at future hadron machines.
For this task, two approaches were proposed. The first was based 
on the QCD factorization estimate of the corresponding penguin parameter 
$(r,\theta)$ assuming the control over the non-perturbative parameters, namely 
the subleading effects. 
The second one, proposed the use of actual experimental 
informations on the $B_d\to \pi^+ \pi^-$ system in order to estimate our 
penguin parameter $(r,\theta)$, using the $SU(3)$-flavour-symmetry argument.
To corroborate the validity of our latter approximation,
we analyzed the $SU(3)$-symmetry-breaking factor $\zeta_{SU(3)}$ within QCD 
factorization with a particular view on theoretical uncertainties.
\vspace*{-0.2cm}
\subsection*{ Acknowledgements:} \vspace*{-0.2cm}
This work is supported by the DFG under contract BU 1391/1-2. 

\vfill\eject

\end{document}